\documentclass[fleqn,usenatbib]{mnras}
\usepackage{fix-cm}

\usepackage{lscape}
\usepackage{float}
\usepackage{subcaption}
\usepackage{multicol}

\usepackage[T1]{fontenc}

\DeclareRobustCommand{\VAN}[3]{#2}
\let\VANthebibliography\thebibliography
\def\thebibliography{\DeclareRobustCommand{\VAN}[3]{##3}\VANthebibliography}

\usepackage{graphicx}	
\usepackage{amsmath}	
\usepackage{amssymb}

\title[Stellar Populations in Satellite Galaxies]{Stellar Populations in Satellite Galaxies in Close Pairs}
\author[A. E. Sansom et al.]{
A. E. Sansom$^{1}$\thanks{E-mail: AESansom@lancashire.ac.uk}, I. Ferreras$^{2,3,4}$, B.F. McDonald$^{1}$
\\
$^{1}$Jeremiah Horrocks Institute, University of Lancashire, Preston, PR1 2HE, UK\\
$^{2}$Instituto de Astrof\'isica de Canarias, V\'ia L\'actea, 38205 La Laguna, Tenerife, Spain\\
$^{3}$Department of Physics and Astronomy, University College London, London WC1E 6BT, UK\\
$^{4}$Departamento de Astrof{\'i}sica, Universidad de La Laguna, E38206 La Laguna, Tenerife, Spain\\}

\date{Submitted for publication in MNRAS. Manuscript version \today}

\pubyear{2025}

\defcitealias{LaBarbera13}{LB13} 

\begin{document}
\label{firstpage}
\pagerange{\pageref{firstpage}--\pageref{lastpage}}
\maketitle

\begin{abstract}
Satellite galaxies that are near to massive primary galaxies in close pairs can have stellar population ages that are more similar to their primaries than expected. This is one way in which close pairs of galaxies show galactic conformity, which is thought to be driven by assembly bias. Such conformity is seen in ages, morphologies and star formation rates in different samples. This paper revisits a high signal-to-noise SDSS spectroscopic sample, by spectral fitting of new stellar population models, to investigate satellite galaxy properties of age, metallicity and $\alpha$-element abundance. We find the clear signature of age conformity, as previously seen, but no clear evidence for conformity in metallicity or abundance ratios. The offsets showing age conformity are not caused by age-metallicity degeneracies. There is a suggestion in these data that lower velocity dispersion satellites have increased [$\alpha$/Fe] compared to a control sample of passive galaxies, however this needs further observations to be verified. Our results also suggest an intriguing turnover in the age trends of the satellites at the highest velocity dispersion, perhaps reflecting the onset of environment-related processes in the most massive groups.
\end{abstract}



\section{Introduction}

Environmental impacts on galaxy evolution have been extensively investigated, in a range of ways and with various findings \citep[e.g.][]{BlantonMoustakas09,Peng10,Kawinwanichakij17}. Environmental effects seem particularly important in the local Universe and less so out to larger redshifts  \citep[e.g.][]{Darvish16,Sattari23}. One way to study such effects in the local Universe is to look at galaxies in groups, using either the most massive galaxy, referred to as the primary galaxy in a group, or the group halo mass as proxies for the environment of other galaxies in those groups \citep{Weinmann06,Pasquali15,Ferreras17}. Primaries are often the central and brightest galaxies within a group, although that is not always the case (\citealt{Skibba11}). Such studies have revealed environmental effects often referred to as 'galactic conformity'. Satellite galaxies are seen to be more similar to their more massive primary galaxies in their morphologies \citep[e.g.][]{Otter20}; star formation \citep[e.g.][and references therein]{Otter20,Mesa21} and in the ages of their stellar populations \citep{LaBarbera14}. These studies show that satellites are more quenched than the general population of galaxies at a given stellar mass.

To study this effect in stellar populations \cite{Ferreras19a} created a sample of stacked Sloan Digital Sky Survey (SDSS) spectra, for satellite galaxies around primary galaxies of different masses. They used stellar velocity dispersion ($\sigma$) as an indicator of the local driver for global galaxy characteristics including their star formation histories, because it is known to correlate well with other galaxy parameters \citep{Rogers10,Ferreras19b,Ferreras:25}. Stellar population models from \cite{BC03} and \cite{Vazdekis12} were fitted to those stacks, using Lick spectral indices (\citealt{Worthey98}), leading to estimates of relative ages and metallicities. From that work \cite{Ferreras19a} found a significant tendency for satellite galaxies around more massive primaries to be older than those around less massive primaries, for a given satellite velocity dispersion. This age difference was particularly evident for satellite galaxies at lower velocity dispersion.

Single Stellar Population (SSP) models, defined as a model of a stellar population with a single age, metallicity and abundance pattern [$\alpha$/Fe], can be used to characterise and compare average properties of galaxies in different subsets (e.g. \citealt{Vazdekis15}). Average $\alpha$-element to iron-peak abundance ratios from integrated light spectra depend on the duration of star formation in a galaxy \citep[see, e.g.][]{Worthey92,MattrRecc:01,FS:02,TMB:05,Martin-Navarro18}. This dependence makes [$\alpha$/Fe] a useful additional indicator of the star formation history in a galaxy, besides average age measurements. In this work we apply new semi-empirical SSP models (\citealt{Knowles23}) that sample a range in [$\alpha$/Fe] to study both age and elemental composition variations within the \cite{Ferreras19a} satellites.

The structure for this paper is as follows. Section~\ref{sec:SDSSsample} describes the galaxy sample used. Section~\ref{sec:SSPmethods} presents details of the spectral fitting and SSP templates used. Section~\ref{sec:SSPresults} shows the findings from these fits, which are then discussed in Section~\ref{sec:Discussion}. Finally Section~\ref{sec:Conclusions} presents our conclusions and discusses future directions.

\section{SDSS Sample}
\label{sec:SDSSsample}
SDSS data are well suited to investigate the stellar populations of galaxies with different stellar masses because of the quality, number and mass range of galaxies with good signal-to-noise spectra. A number of studies with stacked SDSS spectra have indicated effects of environment in the nearby Universe, especially for low and highest mass galaxies (e.g. \citealt{Rogers10}; \citealt{LaBarbera14}). The data used in this current work is based on the sample of satellite galaxy spectra compiled by \citet{Ferreras19a}. Full details are given in that paper, however here we summarise the main points. Satellite and primary galaxies were first selected from SDSS DR14 in the redshift range 0.07 $<$ z $<$ 0.14, with average signal-to-noise ratio per spectral pixel of S/N $>$ 10 in the $r$ band. Primaries have stellar masses $M_* > 10^{11} M_\odot$ with nearby satellites within a projected distance of $\Delta r_\perp\leq 100$ kpc and redshift offsets corresponding to relative velocities within $|\Delta v_\parallel|\leq 700$ km s$^{-1}$ of their primaries. The ratio of satellite mass to primary mass was evaluated for each satellite $\mu = M_{\rm SAT}/M_{\rm PRI}$. The satellites were stacked into five bins in stellar velocity dispersion covering $100 < \sigma < 250\,\rm{km s}^{-1}$ and three bins in mass ratio, $\mu$. Error arrays are generated from the error in the mean in each spectral pixel, for each stack. These restrictions in redshift, signal-to-noise and velocity dispersion were all aimed at making the selection as unbiased as possible, whilst keeping enough galaxies to maintain a high signal-to-noise ratio (S/N$\gtrsim$100 around 5000 \AA) when stacked. This work improves on the previous methodology by implementing an algorithm that removes bad individual pixels from the spectra. After masking out bad pixels from each spectrum, we perform a linear interpolation onto a reference (rest-frame) wavelength 
interval $\lambda\in$[3750,7000]\AA\ in log-$\lambda$ steps of $10^{-4}$, when the wavelength is measured in Angstrom. 
Here we make use of all five bins in velocity dispersion, but concentrate on only the lowest and highest mass ratio bins, namely $0.158 < \mu_3 < 0.324$ and $0.527 < \mu_1 < 1.000$ respectively. Thus, at fixed velocity dispersion, $\mu_3$ represents satellites around more massive primaries, with respect to $\mu_1$. As a control sample, we also produce a set of stacks in the same selection of five bins in velocity dispersion, but including all SDSS galaxy spectra with the same constraint on S/N. The control sample stacks are contrasted with the satellite spectra to assess the significance of the difference between the $\mu_1$ and $\mu_3$ stacks at fixed velocity dispersion.

\begin{figure}
  \includegraphics[width=80mm,angle=0]{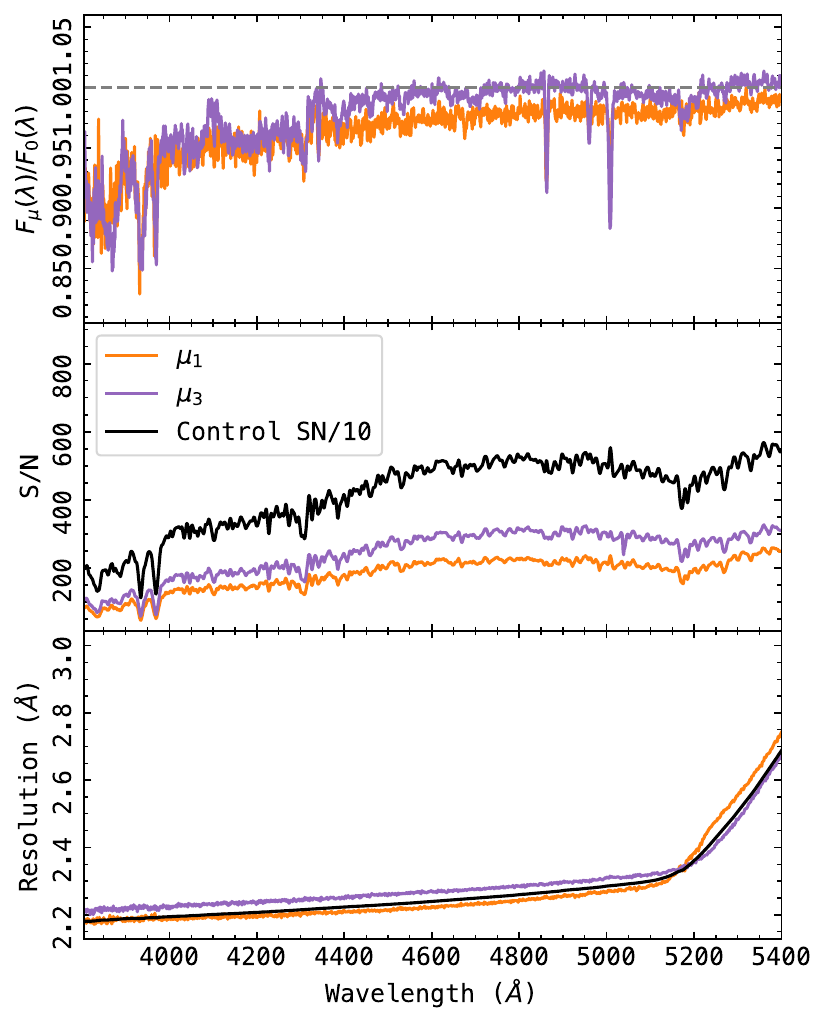}
    \caption{We show some details of the stacks corresponding to the second velocity dispersion bin ($\sigma_2$). The top panel shows the ratio between the close pairs stack ($\mu_1$ in orange and $\mu_3$ in purple) and the corresponding control stack. The middle panel shows the S/N as a function of wavelength, and the bottom panel corresponds to the effective spectral resolution. The characteristic transition in resolution from the blue to the red arm of the SDSS spectrograph is evident \citep{Smee:13}.}
    \label{fig:stack}
\end{figure}

Figure~\ref{fig:stack} shows an example of the stacks, corresponding to $130<\sigma<160$\,km/s.
The top panel shows the ratio between either the $\mu_1$ stack (orange) or the $\mu_3$ stack (purple) with respect to the control sample. Note the excess flux in the NUV of the control sample with respect to either $\mu_1$ or $\mu_3$. A more subtle difference is found between the two close pairs, so that $\mu_1$ shows a shallower wavelength dependence of this flux ratio. The middle and bottom panels show the S/N and the effective spectral resolution, respectively. The S/N is derived from the error in the mean in each spectral bin of every stack. The characteristic variation in S/N and resolution is evident as the data shifts from the blue to the red arm of the SDSS spectrograph. We note that the stacks are studied at their respective spectral resolution, measured by the code {\sc pPXF} \citep{ppxf}, listed in Table~\ref{tab:sample}. We use the standard MILES spectral models \citep{Vazdekis:10} for the {\sc pPXF} fits, although we do not aim at this stage to produce population parameters, leaving that to the next section, where more detailed models are adopted.

\begin{figure}
  \includegraphics[width=80mm,angle=0]{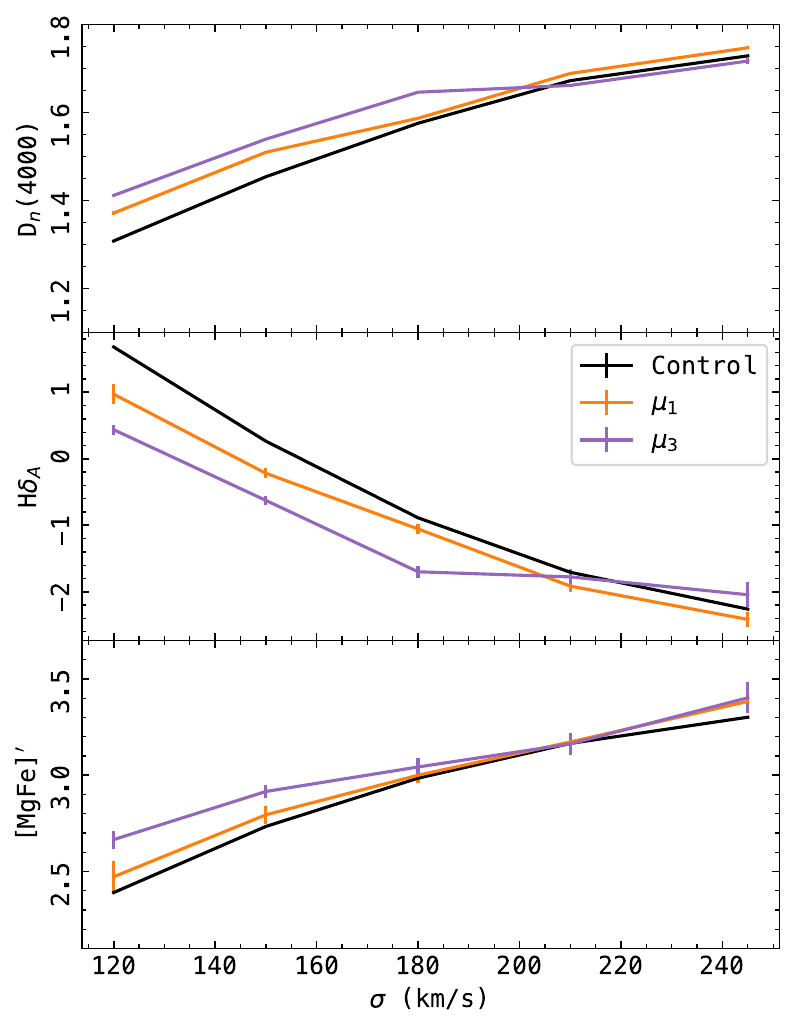}
    \caption{Equivalent widths of the stacks shown as a function of velocity dispersion. From top to bottom we show the 4000\AA\ break strength, D$_n$(4000), Balmer absorption (H$\delta_A$), and the [MgFe]$^\prime$ index. Note the turnover of the age sensitive indices at higher velocity dispersion for the $\mu_3$ stacks. For a consistent comparison, the spectra have been corrected for emission lines and smoothed to a common velocity dispersion of 250\,km/s (see text for details).
    Error bars are propagated from the error in the mean of the stacks, shown at the 1\,$\sigma$ level.}
    \label{fig:EWs}
\end{figure}

As a first estimate of the difference in the stellar population content in these stacks, we show in Figure~\ref{fig:EWs} three typical line strengths as a function of $\sigma$. From top to bottom, the figure shows D$_n$(4000); H$\delta_A$; and [MgFe]$^\prime$. D$_n$(4000) is defined as the ratio of fluxes from 100\AA-wide bands either side of the 4000\AA\ break \citep{Balogh:99}. H$\delta_A$ is the strength of the H$\delta$ Balmer absorption in the wider definition of \citet{WO:97}, and [MgFe]$^\prime\equiv\sqrt{Mgb(0.72\times Fe5270+0.58\times Fe5335)}$ is a composite of three Lick indices designed to minimize the effects of [$\alpha$/Fe] variations \citep{TMB:03}. The Balmer line has been corrected for emission, using the best fit from the {\sc pPXF} run. Moreover, the line strengths in this figure are derived from spectra smoothed to a common velocity dispersion of 250\,km/s. This is produced by convolving the spectra with a Gaussian kernel. {\sc pPXF} is run once more to confirm that the convolved stacks have the target velocity dispersion, within uncertainties. The figure reveals the typical correlation of these indices with $\sigma$, as expected from the well-known age-mass and metallicity-mass relations \citep[e.g.][]{AG:05}. At low $\sigma$, the control sample, in black, has lower 4000\AA\ break strength, lower [MgFe]$^\prime$, and more positive H$\delta_A$, characteristic of younger populations. Within the satellite sample, those around more massive primaries (i.e. $\mu_3$, in purple in the figure) appear consistently older (stronger 4000\AA\ break and less positive H$\delta_A$), in agreement with our previous work \citep{Ferreras17,Ferreras19a}. At the high S/N of the stacked data, it is also worth noting that there is a turnover in the age-sensitive indices D$_n$(4000) and H$\delta_A$ at  $\sigma\gtrsim$200\,km/s, where $\mu_3$ stacks feature lower 4000\AA\ break strength and H$\delta_A$, suggestive of a reversal in the age difference in satellites around more ($\mu_3$) or less ($\mu_1$) massive primaries. This is an important issue that will be explored below with models.

\begin{table*}
   \caption{General properties of the stacks used in the analysis. Each row corresponds to one of the five bins in stellar velocity dispersion. $\mu_1$ and $\mu_3$ represent the two extremes in the stellar mass ratio between satellite and primary. The control sample extends to the whole SDSS Legacy database. For each choice, we show the number of spectra used in each stack, the median S/N per pixel in the rest-frame interval 5,000-5,500\AA (using the error in the mean), and the velocity dispersion ($\sigma$ in km/s) measured by {\sc pPXF} \citep{ppxf}.}
    \centering
    \begin{tabular}{cc|ccc|ccc|ccc}
    \hline
Vel.Disp. & Range & \multicolumn{3}{c}{$\mu_1$} & \multicolumn{3}{c}{$\mu_3$} & \multicolumn{3}{c}{Control}\\
 bin &  km/s  & N & S/N & $\sigma$ & N & S/N & $\sigma$ & N & S/N & $\sigma$\\ 
        \hline
 $\sigma_1$ & 100-130 &  36 & 101 & 120.0$\pm$2.6 & 188 & 229 & 119.7$\pm$1.5 & 71593 & 4383 & 119.5$\pm$1.5\\
 $\sigma_2$ & 130-160 & 140 & 215 & 154.0$\pm$1.6 & 256 & 291 & 149.6$\pm$1.4 & 69379 & 4947 & 149.2$\pm$1.3\\
 $\sigma_3$ & 160-190 & 205 & 286 & 181.0$\pm$1.5 & 128 & 219 & 178.2$\pm$1.6 & 51734 & 4768 & 179.7$\pm$1.3\\
 $\sigma_4$ & 190-220 & 172 & 287 & 213.0$\pm$1.7 &  70 & 168 & 209.6$\pm$2.1 & 29165 & 3960 & 211.8$\pm$1.5\\
 $\sigma_5$ & 220-250 &  85 & 223 & 243.6$\pm$2.1 &  23 & 100 & 241.8$\pm$2.7 & 13973 & 3052 & 245.3$\pm$1.8\\
 \hline
    \end{tabular}
    \label{tab:sample}
\end{table*}

\section{Measuring Stellar Population Parameters}
\label{sec:SSPmethods}
In this current work we look at three stellar population parameters simultaneously. These are light weighted mean stellar population age, metallicity and [$\alpha$/Fe] ratio, where the $\alpha$-element dependence is dominated by magnesium sensitivity in our approach. Fitting all three parameters with SSPs simultaneously becomes possible with our new SSP models that are described below. 

\subsection{New spectral templates} 
\label{SSPmodels}
The new SSP model spectra that we use in this work are from \cite{Knowles23} and cover a wide range in ages (0.03 to 14 Gyr), total metallicities ($-1.79 \le [M/H] \le +0.26$ dex\footnote{Labelled as $[M/H]_{SSP}$ in \citealt{Knowles23}} and also 5 steps in [$\alpha$/Fe] of -0.2, 0.0, +0.2, +0.4 and +0.6 abundance ratio. Most previous models only covered two steps in [$\alpha$/Fe] for the full spectrum (e.g. \citealt{Vazdekis15}). These new models are built from semi-empirical stars made by combining empirical MILES library stars\footnote{https://research.iac.es/proyecto/miles/} with modifications from theoretical star spectra\footnote{https://data.lancashire.ac.uk/178/} that have variable abundances of $\alpha$-capture to iron peak elements. These SSP models and the semi-empirical stars from which they were made are described in detail in \cite{Knowles23} and \cite{Knowles21} respectively. The theoretical star spectra that went into making the semi-empirical star spectra have $\alpha$-elements O, Ne, Mg, Si, S, Ca and Ti all varying together both in the photospheres and in the radiative transfer modelling of individual stars. All other elements were varied together scaled to iron. This means that the [$\alpha$/Fe] ratios were handled self-consistently throughout the generation of theoretical star spectra. Thus there was no need to assume that the $\alpha$-elements group were behaving only as trace elements (as is assumed by \citealt{Conroy18}). This self-consistency of approach propagates through to the SSPs, at the expense of being able to follow multiple elements individually (e.g. as in \citealt{Conroy18}). Evidence from stars in our own Galaxy (e.g. \citealt{Bensby14}; \citealt{Weinberg19}) indicates that the $\alpha$-elements vary approximately as a group, although there are some known smaller systematic differences. For example, for low metallicity disc stars in our Galaxy [O/Fe] is more enhanced than [Mg/Fe], which in turn is more enhanced than heavier $\alpha$-capture elements such as Si, Ca, Ti (e.g. see \citealt{Bensby14} their figure 15; \citealt{Hourihane23} their figure 23).

\subsection{SSP fitting methodology}
\label{sec:SSPfitting}
To investigate stellar populations in satellite galaxies and how they might be influenced by their location near to primaries, we look at how the satellite galaxy populations compare in their light weighted mean ages, metallicities and [$\alpha$/Fe] ratios, at a given satellite velocity dispersion, for the low and high mass ratio categories described above. \small{PYTHON} \normalsize{software} was written to use the new SSP libraries of \cite{Knowles23} to search for best fitting SSPs and uncertainties in their three parameters. This SSP fitting software is described in \cite{Knowles23}. In this work, 15 Lick line indices in the optical range are fitted, to spectra at fixed $\sigma$ for each stack, by blurring SSP spectra to the central velocity dispersion of each galaxy stack. The Lick line indices included are: H$\delta_{A\&F}$, G4300, H$\gamma_{A\&F}$, Fe4383, Ca4455, Fe4531, H$\beta$, Mgb, Fe5270, Fe5335, Fe5406, Fe5709 and Fe5782 \citep[defined in][]{WO:97,Trager:98}. 
Amongst these indices the sensitivity to $\alpha$-elements is mainly through the magnesium sensitivity of Mgb. Hence our fits are more sensitive to [Mg/Fe] than global [$\alpha$/Fe] changes. We use Lick indices, rather than full spectrum fitting, to include the spectral features that are most sensitive to the stellar population parameters that we aim to measure. This is important because there appear to be some problems in fitting full spectra with $\alpha$-enhancement (\citealt{Pernet24}). In this work a fixed bimodal initial mass function {\citep[comparable to][]{Ku:01}} is assumed, which is reasonable given the lack of sensitivity to the stellar initial mass function for galaxies with velocity dispersions below $\sim 250$ km s$^{-1}$ (\citealt{Pernet24}, see their figure 4). 

To derive error estimates for the three stellar population parameters, Monte Carlo tests are used, varying the spectral fluxes in the satellite stacks by their uncertainties, Gaussian sampled assuming the Gaussian 1-sigma is the uncertainty in each spectral bin of the stacks.  

\subsection{Application to SDSS satellite galaxy stacks}
\label{sec:Application}
Fitting SSPs to galaxy populations is a useful, controlled way of looking at {\it relative} behaviours in derived average population parameters (e.g. \citealt{TMB:05}; \citealt{LaBarbera13}; \citealt{Pernet24}), without introducing many free parameters. However, galaxy spectra are more complex than this and models are imperfect, therefore the spectral fits to high S/N spectra are imperfect, and hence we note that absolute values in derived average population parameters are less certain. In this work we concentrate on looking at relative behaviours between satellite galaxies in different environments, focusing on SSP-equivalent results. 

\begin{figure}
\vspace{-1.5cm}
\includegraphics[width=92mm]{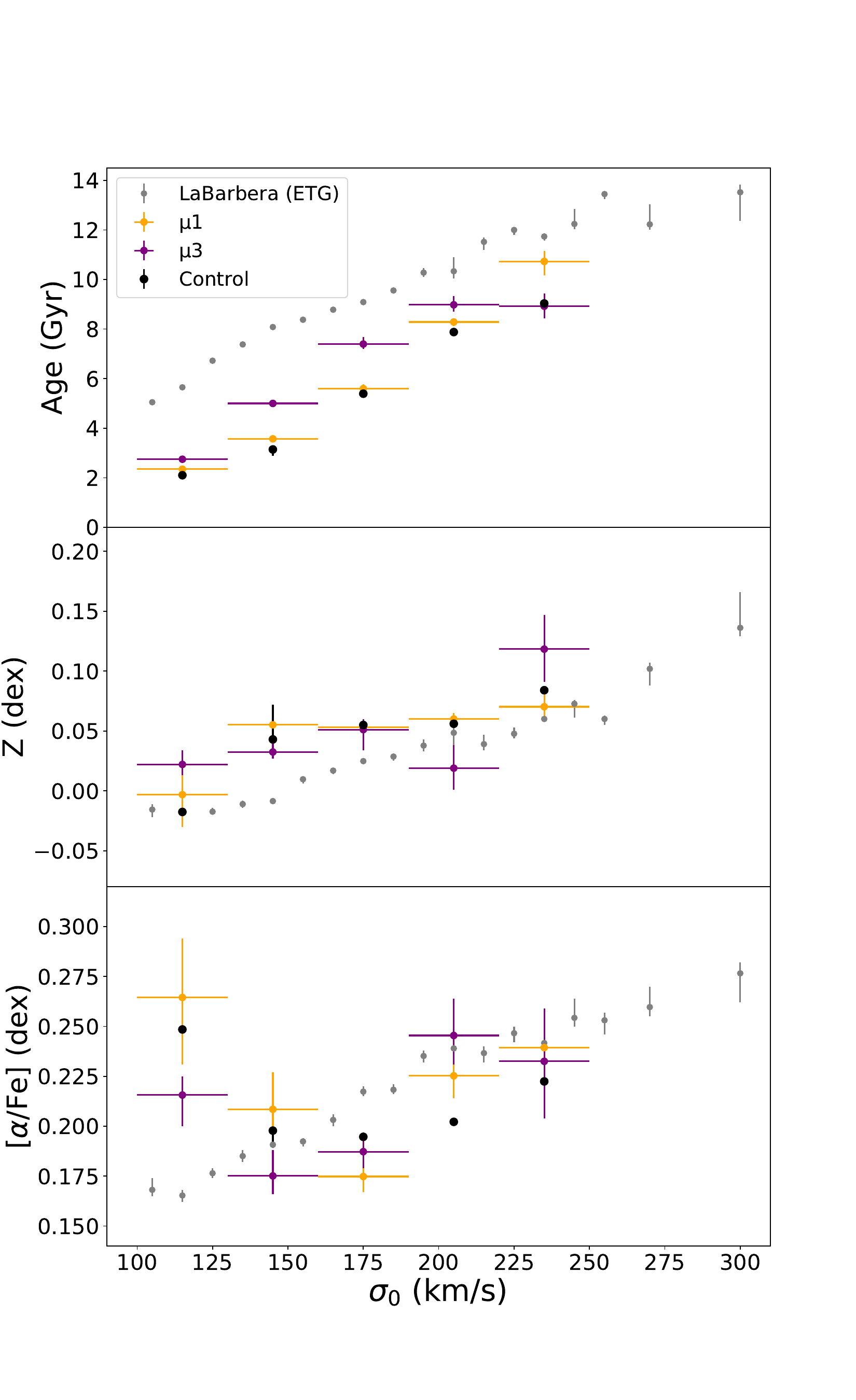}
\caption{SSP fits to SDSS stacked satellite galaxy data (from \citealt{Ferreras19a}). Light-weighted ages (top panel), metallicities Z=[M/H]$_{SSP}$ (middle panel) and [$\alpha$/Fe] ratios (lower panel) are given, based on fitting 15 Lick line indices, plotted against stellar velocity dispersion in the middle of each bin ($\sigma_0$). Orange points are for satellites with masses just below their nearby primary galaxies ($\mu_1$); purple points are for satellites with much lower masses than their primaries ($\mu_3$); black points are for the control sample without environment constraints; small grey points are from the sample of passive, early-type galaxies (ETG) described in Section~\ref{sec:Full}. Uncertainty ranges are obtained from our Monte-Carlo perturbations of the spectra being fitted (see Section~\ref{sec:SSPfitting}). 
}
  \label{SSP_Params_v_Sigma}
\end{figure}

We fit the stacks initially and then use those initially fit SSP models to investigate if there is any residual emission line flux in the galaxy spectra. This revealed some weak excess flux ($<3\%$) at the expected locations of some emission lines that could affect the indices being measured. Therefore, we isolate residual emission lines over small wavelength ranges ($\sim 20$ \AA) and replace the flux by this initial best fit SSP model, to minimize the effects of line emission in the galaxy spectra. Then fitting is carried out a second time, on these emission line removed spectra, leading to the best fits for light weighted average stellar population parameters from the satellite galaxy stacked spectra. Best fits, errors from perturbations and scaled $\chi^2_{\nu}$ values for the best fits are given in Table~\ref{SSPfits_Table} and parameters are plotted in Figure~\ref{SSP_Params_v_Sigma}. The fits are shown in this table for $\mu_1$ and $\mu_3$ stacks and their differences $(\mu_3 - \mu_1)$, for the three SSP parameters fitted. 
The $\chi^2_{\nu}$ values only take into account errors in mean spectral values in the stacks and do not account for unknown errors in the model templates or differences between galaxy spectra within a stack. Those additional errors are difficult to quantify. We attempt to consider them by scaling the  $\chi^2_{\nu}$ values. Therefore, the absolute values derived in these fits are not so reliable, but this work concentrates on looking for relative differences between stacks. This scaling is discussed further in Appendix~\ref{app:c2nu}.


\begin{table*}
   \caption{ 
   Stellar population parameters for best fits (minimizing $\chi^2_{\nu}$) to emission-line subtracted stacks of SDSS satellite galaxy spectra, by fitting 15 Lick indices (see Section~\ref{sec:SSPfitting}). The first column shows the velocity dispersion ($\sigma$) bin range being measured. The second column indicates which data set is being measured, in term of mass ratio $\mu=M_{\rm SAT}/M_{\rm PRI}$, or Control sample. The third, fourth and fifth columns give the best fitting Age, metallicity and [$\alpha$/Fe] values together with their lower and upper ranges from flux density perturbation error analysis (see Section~\ref{sec:SSPfitting}). The best fit $\chi^2_{\nu}/{\sqrt{N_{g}}}$ is given, where $N_{g}$ is the number of galaxies in the stack (see Appendix~\ref{app:c2nu} for details).
   For each $\sigma$ bin, there are four rows: Row 1 gives results for those satellites that have a similar mass to their nearby primaries ($\mu_1$); Row 2 gives results for those satellites that are much less massive ($\mu_3$) than their primaries; 
   Row 3 gives results for the Control sample. Row 4 gives the difference between $\mu_3$ and  $\mu_1$ parameters.
   Recall that all primaries considered in this work are massive ($M_* > 10^{11} M_\odot$), as detailed in Section~\ref{sec:SDSSsample}.}
    \centering
    \begin{tabular}{|c|c|p{32mm}|p{32mm}|p{32mm}|c|}
    \hline
    Vel. Disp. & Mass ratio bin & Age & [M/H]$_{\textrm{SSP}}$ & [$\alpha$/Fe] & $\chi^2_{\nu}/\sqrt{N_{g}}$  \\
    bin &  & (Gyr) & (dex) & (dex) & \\
        \hline
    $\sigma_1$  & $\mu_1$ & 2.346 ({2.255 to 2.453}) & $-$0.003 ({-0.030 to 0.024}) & $+$0.264 ({0.231 to 0.294}) & 1.53  \\
                & $\mu_3$ & 2.748 ({2.678 to 2.750}) & $+$0.022 ({0.013 to 0.034}) & $+$0.216 ({0.200 to 0.225}) & 1.05 \\
                & Control & 2.103 ({2.102 to 2.106}) & $-$0.018 ({-0.019 to -0.017}) & $+$0.248 ({0.248 to 0.250}) & \\
                & $\Delta=\mu_3-\mu_1$ & 0.402 & 0.025 & $-$0.048 &  \\
               & & & & & \\           
    $\sigma_2$ & $\mu_1$ & 3.571 ({3.500 to 3.699}) & $+$0.055 ({0.041 to 0.060}) & $+$0.208 ({0.197 to 0.227}) & 1.28  \\
                & $\mu_3$ & 5.000 ({4.877 to 5.000}) & $+$0.032 ({0.027 to 0.056}) & $+$0.175 ({0.166 to 0.188}) & 1.26 \\
                & Control & 3.143 ({2.887 to 3.146}) & $+$0.043 ({0.042 to 0.072}) & $+$0.198 ({0.192 to 0.198}) & \\
                & $\Delta=\mu_3-\mu_1$ & 1.429 & $-$0.023 & $-$0.033 &  \\
                & & & & & \\
    $\sigma_3$ & $\mu_1$ & 5.598 ({5.500 to 5.768}) & $+$0.053 ({0.040 to 0.060}) & $+$0.175 ({0.167 to 0.183}) & 1.52  \\
                & $\mu_3$ & 7.395 ({7.203 to 7.685}) & $+$0.051 ({0.034 to 0.060}) & $+$0.187 ({0.179 to 0.197 }) & 1.23 \\
                & Control & 5.392 ({5.371 to 5.406}) & $+$0.055 ({0.054 to 0.057}) & $+$0.195 ({0.194 to 0.195}) & \\
                & $\Delta=\mu_3-\mu_1$ & 1.797 & $-$0.002 & 0.012 &  \\
                & & & & & \\
    $\sigma_4$  & $\mu_1$ & 8.286 ({8.113 to 8.435}) & $+$0.060 ({0.056 to 0.065}) & $+$0.225 ({0.214 to 0.235}) & 1.82 \\
                & $\mu_3$ & 8.979 ({8.698 to 9.336}) & $+$0.019 ({0.001 to 0.038}) & $+$0.245 ({0.231 to 0.264}) & 1.22 \\
                & Control & 7.879 ({7.868 to 7.902}) & $+$0.056 ({0.055 to 0.057}) & $+$0.202 ({0.202 to 0.203}) & \\
                & $\Delta=\mu_3-\mu_1$ & 0.693 & $-$0.041 & 0.020 &  \\
                & & & & & \\
    $\sigma_5$  & $\mu_1$ & 10.729 ({10.17 to 11.15}) & $+$0.070 ({0.060 to 0.082}) & $+$0.239 ({0.225 to 0.252}) & 1.84  \\
                & $\mu_3$ & 8.921 ({8.428 to 9.427}) & $+$0.118 ({0.091 to 0.147}) & $+$0.232 ({0.204 to 0.259}) & 1.02 \\
                & Control & 9.034 ({9.009 to 9.064}) & $+$0.084 ({0.083 to 0.085}) & $+$0.222 ({0.221 to 0.223}) & \\
                & $\Delta=\mu_3-\mu_1$ & $-$1.808 & 0.048 & $-$0.007 &  \\
                & & & & & \\
    \end{tabular}
    \label{SSPfits_Table}
\end{table*}

\section{Results}
\label{sec:SSPresults}
This section describes the results of the SSP fits to stacked spectra of satellite galaxies.

\subsection{Full Satellite Stacks}
\label{sec:Full}
Using the full satellite stacks described above we measured their stellar population parameters. In Figure~\ref{SSP_Params_v_Sigma}, we show how the best fitting parameters compare, for different $\mu=M_{\rm SAT}/M_{\rm PRI}$ at the lowest ($\mu_3$) and highest ($\mu_1$) mass ratios, plus in the control sample. Broadly, this figure shows that stellar populations increase in age with increasing velocity dispersion (as is well known from many other studies, e.g. \citealt{LaBarbera13}; \citealt{McDermid15}; \citealt{Ferreras19b}).

Also plotted in Figure~\ref{SSP_Params_v_Sigma} are the trends that we found in \cite{Knowles23} for SDSS data of passive galaxies from \cite{LaBarbera13}, which we use here as a comparison sample of quiescent galaxies. All points plotted in Figure~\ref{SSP_Params_v_Sigma} are measured in the same way, using our SSP fitting \small{PYTHON} \normalsize{code}. The grey points of the quiescent sample show age, metallicity and [$\alpha$/Fe] all increasing with $\sigma$. The ages in our satellites are generally younger than in the quiescent sample but follow a similar slope. Note that our satellite galaxies are not selected to be passive. Metallicities are estimated as slightly higher, and with greater uncertainty in the satellites than in the quiescent sample. There may be some age-metallicity degeneracy systematic affecting the satellites samples, but this is not clear in the trends observed (offsets from the quiescent sample do not anticorrelate). The [$\alpha$/Fe] values are similar at higher $\sigma$, but differ from the quiescent sample in the lowest $\sigma$ bin.

Whilst age increases with $\sigma$ for both $\mu_1$ and $\mu_3$ satellite stacks, we see an offset between these two (orange and purple points respectively). The top panel in Figure~\ref{SSP_Params_v_Sigma} highlights systematic differences in ages of stellar populations, with low mass satellites around massive primaries ($\mu_3$) showing older ages than in satellites with more similar masses to their massive primaries ($\mu_1$) at a given satellite galaxy $\sigma$. This difference in stellar population ages, for satellite galaxies with the same $\sigma$ highlights the impact of their environments, through being close to a primary galaxy. If there was no such environmental influence then there would be no difference between galaxies at the same $\sigma$. This is a manifestation of conformity and it agrees qualitatively with what was found in \cite{Ferreras19a}. The effect on ages is at the level of up to $\sim 2$\,Gyr difference. The control sample accentuates this difference because it shows younger ages at a given $\sigma$ than either of the satellite stacks. Note that the satellite stacks show a reversal of behaviour in ages for the highest $\sigma$ bin, similar to the reversal seen in age-sensitive indices in Figure~\ref{fig:EWs}.

The middle panel in Figure~\ref{SSP_Params_v_Sigma} indicates no significant systematic difference between $\mu_1$ and $\mu_3$ classes of satellite galaxies, at a given $\sigma$, hence we do not detect any systematic effect on metallicity of stellar populations in satellite galaxies resulting from proximity to a massive primary galaxy. This lack of systematic difference may indicate that higher signal-to-noise is needed in the spectra to see such effects, or that light weighted metallicity is less sensitive to environmental impact than light weighted age. Because metallicity is generally determined by redder wavelength features than age, in a composite population, then light-weighted metallicity estimates are more influenced by older stars than are the light-weighted age estimate. Satellites may experience a range of environments through their histories as they are pre-processed in field or group environments before their their current group is formed (\citealt{Oxland24} and references therein). Larger scale group and cluster formation occurs later than the formation of galaxies within those groups, therefore the metallicities may be more strongly affected by their histories, whereas ages are more strongly affected by the most recent star formation. Their evolution in these enviroments will determine their average stellar abundances (e.g. \citealt{Bidaran22}). These effects, and the uncertainties in measuring metallicities may contribute to why we do not see a systematic effect for metallicity. The control sample is also consistent with metallicity estimates for the satellites.

The lowest panel in Figure~\ref{SSP_Params_v_Sigma} indicates that [$\alpha$/Fe] values tend to increase towards higher $\sigma$, as is observed generally for more massive, passive galaxies (e.g. \citealt{LaBarbera13}; \citealt{Martin-Navarro18}). However, there is a suggestion of an upturn in [$\alpha$/Fe] for our lowest $\sigma$ bin ($\sim 115$\,km s$^{-1}$, for all samples of our stacks ($\mu_1$, $\mu_3$ and control samples). The quiescent sample does not show such an upturn. This upturn needs further investigation in larger samples of galaxies to find out if it is real. Between the $\mu_3$ and $\mu_1$ mass ratio bins there is no evidence for a systematic change in [$\alpha$/Fe], at a given $\sigma$, and hence no evidence of environment affecting that parameter in satellite galaxies in close pairs with their primary galaxies.

\begin{figure*}
\includegraphics[width=\linewidth]{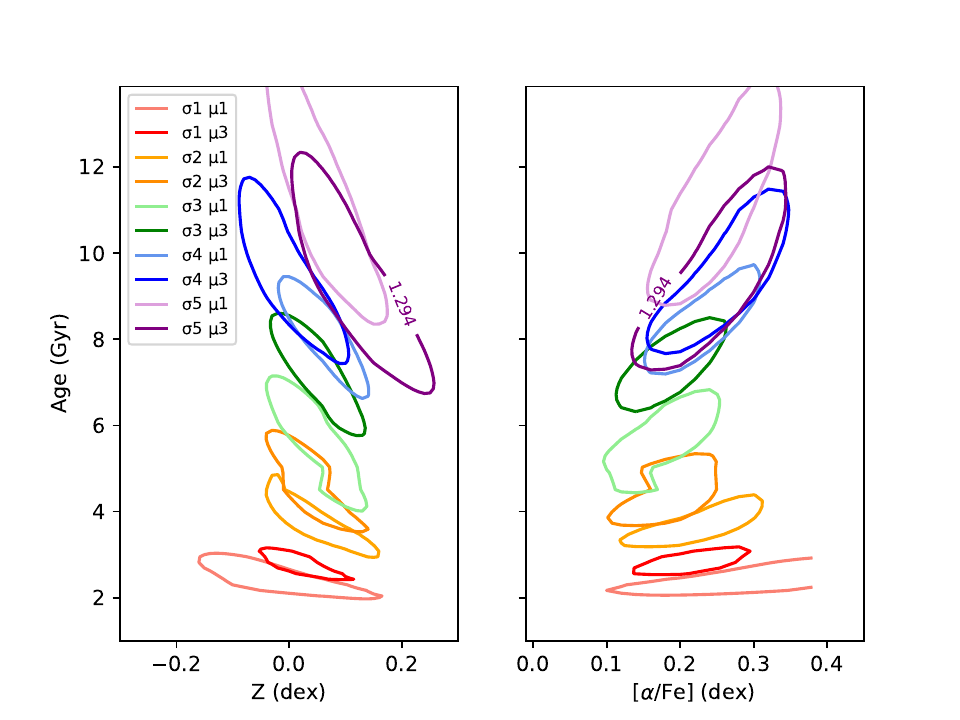}
\caption{This figure shows normalised $\chi^2_{\nu}$ contours for SSP fits to the satellite galaxy stacks at different stellar velocity dispersions, indicated by the five different colours. Contours are plotted at 1-sigma confidence level for three fitted parameters, fitting 15 Lick indices. Here $\sigma_1$ to $\sigma_5$ are the lowest to highest velocity dispersion bins as listed in Table~\ref{SSPfits_Table}. At each $\sigma$, satellite stacks with low $M_{SAT}/M_{PRI}$ ($\mu_3$) are shown with a darker contour line and satellite stacks with higher $M_{SAT}/M_{PRI}$ ($\mu_1$) are shown with a lighter contour line. On the left are plotted contours in Age and metallicity ($=Z=$[M/H]$_{SSP}$) space, at the best fitting [$\alpha$/Fe]. On the right are plotted contours in Age and [$\alpha$/Fe], at the best fitting metallicity.} 
  \label{SSP_Contours}
\end{figure*}

To investigate these trends for correlated effects we plot $\chi^2_{\nu}$ contours in two parameters, whilst fixing the third parameter at the best fit value. In Figure~\ref{SSP_Contours} we plot $\chi^2_{\nu}$ contours from grids of our SSP fits, which are also generated by the \small{PYTHON} \normalsize{SSP} fitting code. The $\chi^2_{\nu}$ values plotted are normalised by the best fitting model $\chi^2_{\nu}$, to account for galaxy-to-galaxy scatter and systematic differences between the models and data. These normalised $\chi^2_{\nu}$ contour plots allow us to map the distribution of where the best fits occur and to look at results more independently of the well known age-metallicity degeneracy effect. The contours are plotted at a level of $1.0+3.53/(15-3)=1.294$. The value of 3.53 is for 68$\%$ confidence with three free parameters \citep[see, e.g.,][]{Press:92} and there are 15 Lick indices being fitted with 3 parameters.

Figure~\ref{SSP_Contours} (left panel) highlights the systematic difference between $\mu_3$ (darker contours) and $\mu_1$ (lighter contours) satellites in their ages, with $\mu_1$ contours sitting below $\mu_3$ contours, particularly for the lowest $\sigma$ stacks. Whilst metallicity is not very well constrained the offsets between contours in the vertical (age) axis reveal age differences, with relatively low mass ratio satellites ($\mu_3$) appearing older than higher mass ratio  satellites ($\mu_1$). Figure~\ref{SSP_Contours} (right panel) again shows this age offset (vertical axis) but illustrates no significant offset in [$\alpha$/Fe] (horizontal axis).

\subsection{Marginalised Results}
\begin{figure*}
    \includegraphics[width=\linewidth,angle=0]{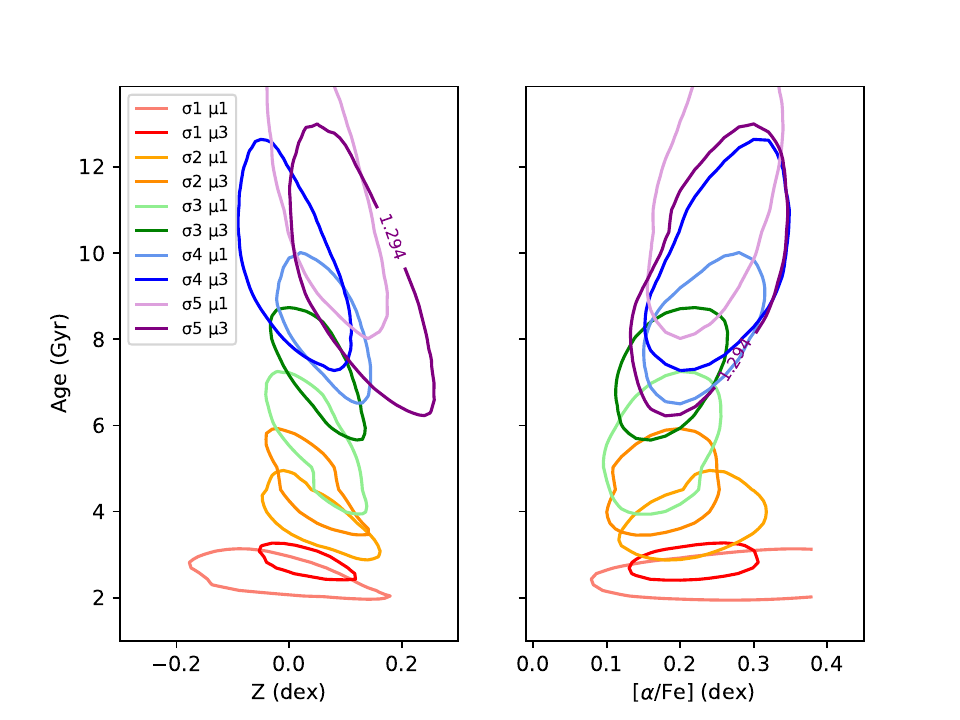}
    \caption{Same as in Figure~\ref{SSP_Contours} but for SSP fits marginalised over the third parameter not plotted in each case.}
    \label{SSP_Contours_Marginalised}
\end{figure*}

For the full satellite stacks we plot the contours again but this time marginalising over the third parameter in each case, as a more stringent test of the offsets. Figure~\ref{SSP_Contours_Marginalised} shows that the age offsets are still present, even when the third parameter (not plotted) is marginalised over. That is, any value of the third parameter is allowed over the input range for that parameter. This equates to collapsing the third parameter axis, rather than fixing the third parameter to its best fit value (as was done in Figure~\ref{SSP_Contours}).

Thus our finding of age offsets in satellite galaxies of the same $\sigma$, due to conformity, remains when we use the new stellar population models with a wide range in [$\alpha$/Fe] values. This result is independent of any age-metallicity degeneracy effect, as seen in Figure~\ref{SSP_Contours} and Figure~\ref{SSP_Contours_Marginalised} left panels.

\section{Discussion}
\label{sec:Discussion}
More massive galaxies tend to have higher velocity dispersions and older  stellar populations. The primary galaxies used to characterise the local environments in groups in this study are massive galaxies ($M_* > 10^{11} M_\odot$). We found here and in \cite{Ferreras19a} and \cite{Ferreras17} that ages of satellite galaxies tend to be more similar to that of their primary galaxies in close pairs than expected for the velocity dispersion of the satellites. This is seen as older ages in satellites with lower masses compared to ther primaries, thus conforming more to the primary galaxy properties than expected. If there was no such conformity effect then, at a given $\sigma$, satellites would not show such systematic offsets in age (Figure~\ref{SSP_Contours_Marginalised}). In this paper we show that this age offset is still present even when [$\alpha$/Fe] is allowed to vary. Age conformity may be a result of cessation of star formation being influenced by the environment of a satellite galaxy, including its proximity to a more massive primary galaxy. For example, ram pressure stripping and starvation of interstellar medium gas may be more effective close to a massive primary galaxy in a group. Then without the gas to form new stars the satellite galaxy becomes passive earlier. The light-weighted average age of the satellite will then depend on when it fell into the group, with more recent acquisitions still possessing younger stars.

This work shows no significant offsets in metallicity between our $\mu_1$ and $\mu_3$ satellite stacks. For the metallicity, this may have been determined mainly by the evolution of the satellite in field and groups prior to its infall into the group of the central. This pre-processing could explain why we do not see effects of conformity in the metallicity  between satellites and centrals. Whilst there is strong evidence of pre-processing of galaxies from observations (\citealt{Hou14}; \citealt{Bidaran22}) and models (\citealt{McGee09}), the exact effects on element abundances, how that leads to scaling relations and offsets, is not well known. More work is needed to be able to statistically follow how stellar population parameters behave in pre-processing and due to current environment effects.

A well studied possibility is that the primaries and secondaries were made at about the same time, with lower mass ratios ($\mu_3$) surviving in older groups (assembly bias, see discussion in \citealt{Ferreras19a} and references therein; \citealt{Wang22}). Assembly bias was revealed and is supported by cosmological simulations (e.g. \citealt{Gao05}; \citealt{Montero-Dorta21}). Is it possible to distinguish between different mechanisms? Changes with redshift or cluster concentration in simulations and comparisons with data may help to distinguish between various scenarios. New data are becoming available in ongoing and future large scale surveys, such as the Dark Energy Spectroscopic Instrument (DESI) survey \citep{DESI} or the WHT Enhanced Area Velocity Explorer (WEAVE) stellar population survey \citep{WEAVE}.

Finally, we also discuss the turnover of the age trend at the highest velocity dispersion. In Figure~\ref{fig:EWs} the ordering of the line strengths with respect to the mass ratio, $\mu$, are found to change in the $\sigma_5$ bin, a result that is reflected in Figure~\ref{SSP_Params_v_Sigma}, so that the $\mu_3$ stacks appear {\sl younger} than $\mu_1$ galaxies, in contrast with all the other stacks at lower velocity dispersion. While the error bars in metallicity and [$\alpha$/Fe] are larger, it is also worth pointing out that this turnover also suggests $\mu_3$ galaxies in the $\sigma_5$ bin having higher metallicity and perhaps lower [$\alpha$/Fe], although this result is rather weak. This result can be interpreted as an external channel of evolution in close pairs, at the highest velocity dispersion, $\mu_3$ galaxies -- that have a very massive primary --  potentially the central galaxy of a massive group or cluster. In such pairs, we would expect the satellite to be affected by the well-known environment effects that produce younger ages in central galaxies of massive groups \citep[e.g.,][]{LaBarbera14,Pasquali15}. In contrast, at lower velocity dispersion, the trends are, rather, suggestive of a more generic effect related to assembly bias, instead of the microphysics of galaxy formation in massive groups. This turnover appears to be rather robust, given the consistent trend in such a large number of high quality SDSS spectra. Therefore, we propose it as a useful test of the subgrid physics in cosmological simulations of galaxy formation \citep[e.g.][]{EAGLE,TNG}.

\section{Conclusions}
\label{sec:Conclusions}
We have used new stellar population models \citep{Knowles23}, with a wider range of [$\alpha$/Fe] values than previously available, to simultaneously fit light-weighted ages, metallicities and [$\alpha$/Fe] values to stacked SDSS spectra of satellite galaxies in close pairs with primary galaxies. With these new SSP models we confirm the results of \cite{Ferreras19a} that satellites with low mass ratios compared to their primaries are systematically older than satellites with masses more similar to their primaries. Figure~\ref{SSP_Params_v_Sigma} and Figure~\ref{SSP_Contours_Marginalised} show this systematic offset for four out of five bins in satellite velocity dispersion. Thus we detect evidence of conformity in ages of satellite galaxies, such that they are more like their primaries than expected for their stellar velocity dispersion. A control sample of galaxies, not selected by environment, is slightly younger than either satellite sample at a given $\sigma$  (Figure~\ref{SSP_Params_v_Sigma}), which is expected because they are not generally close to massive galaxies and hence are not so affected by conformity.

In this work we also looked for evidence of conformity in element abundances. In contrast to ages, we do not detect any systematic effects in the abundances (metallicity or [$\alpha$/Fe]) between low and high mass ratio satellites. Offsets between these mass ratios do not show systematic effects in Figure~\ref{SSP_Params_v_Sigma} or Figure~\ref{SSP_Contours_Marginalised}. Evidence such as that shown here will help to constrain simulation of hierarchical structure and galaxy formation (e.g. \citealt{Lin22}).

Compared to a quiescent sample of passive galaxies (not selected by environment) the satellite galaxies studied here have younger ages (Figure~\ref{SSP_Params_v_Sigma}, top panel); similar metallicities (Figure~\ref{SSP_Params_v_Sigma}, middle panel) and similar [$\alpha$/Fe] except in the lowest $\sigma$ bin (Figure~\ref{SSP_Params_v_Sigma}, lowest panel). Results for the high [$\alpha$/Fe] measurement in satellites in the lowest $\sigma$ bin need more data to confirm that result. 

We also find a robust turnover in the ages of the satellites at the highest velocity dispersion, with $\mu_3$ stacks appearing younger than $\mu_1$ galaxies, a trend that suggests the effect of environment processes in the most massive groups, a result that is potentially a useful test of the baryon physics in numerical simulations.

\section*{Acknowledgements}
AES acknowledges support from the University of Central Lancashire, Jeremiah Horrocks Institute for undergraduate research internship funding for Benjamin McDonald during summer 2023. Thanks to the referee (Prof Jon Loveday) for comments that helped to improve this paper.

\section*{Data Availability}
The semi-empirical SSP libraries used in this work were presented in \cite{Knowles23} and are publicly available on the Lancashire Online Research Data repository ({\url{https://data.lancashire.ac.uk/388/}}) and MILES website ({\url{http://research.iac.es/proyecto/miles/pages/other-predictionsdata.php}}). 




\bibliographystyle{mnras}
\bibliography{AES_SDSS_Satellites_References_r1}




\appendix

\section{Fitting statistic for stacked data}
\label{app:c2nu}

\begin{figure}
    \includegraphics[width=\linewidth,angle=0]{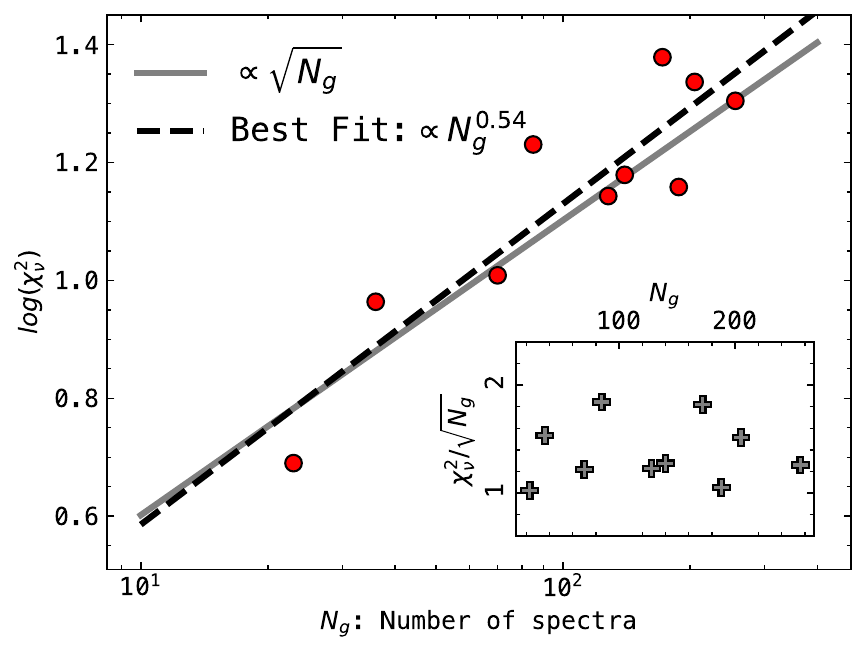}
    \caption{Trend of the reduced $\chi^2$ statistic ($\chi^2_\nu$, derived from the error in the mean) for the SSP analysis, shown with respect to the number of spectra in each stack, $N_g$, along with the best fit (black dashed line), and a simple $N_g^{1/2}$ scaling (grey solid line). The inset shows the effective $\chi^2_\nu$ after being divided by this scaling factor.}
    \label{figA:c2nu}
\end{figure}

Our analysis of stacked data faces an important issue regarding the 
absolute value of the best-fit statistic, i.e. the minimum value of 
the reduced $\chi^2$ statistic ($\chi^2_\nu$). As the number of individual
spectra in each stack is fairly high (see Table~\ref{tab:sample}), the associated uncertainty per spectral bin is rather small, and the resulting $\chi^2_\nu$ is high. However,
in the stacking procedure we typically quote the error in the mean, which would
be an accurate description if each individual spectra were to represent noisy 
realisations of the same galaxy. The stacking of different galaxies implies that, 
in addition to the error in the mean, one should also take into account the 
scatter among galaxies. We follow an empirical approach to assess this 
term by plotting the reduced $\chi^2$ with respect to the number of spectra in each 
bin, $N_g$ (see Figure~\ref{figA:c2nu}). Note the clear trend of the data with respect to a $\sqrt{N_g}$ scaling (grey solid line). A least squares fit (dashed black line) gives a very similar result. Therefore, we adopt the $\sqrt{N_g}$ trend to produce an effective $\chi^2_\nu$ statistic, as quoted in Table~\ref{SSPfits_Table}. Similar types of scaling have been implemented on the best-fit statistic in other studies for a more meaningful comparison of the parameters \citep[see, e.g.][]{Mitzkus:17,Davis:20}. This scaling does not change the best fitting SSP parameter values.

\bsp	
\label{lastpage}
\end{document}